\documentclass[conference]{IEEEtran}
\IEEEoverridecommandlockouts
\usepackage[T1]{fontenc}

\usepackage{cite}
\usepackage{amsmath,amssymb,amsfonts}
\usepackage{algorithmic}
\usepackage{graphicx}
\usepackage{textcomp}
\usepackage{xcolor}

\def\bn{\mbox{\boldmath $n$}}

\def\by{\mbox{\boldmath $y$}}

\def\bs{\mbox{\boldmath $s$}}

\def\by{\mbox{\boldmath $y$}}
\def\mB{\mbox{$\mathbf{B}$}}

\def\mL{\mbox{$\mathbf{L}$}}

\def\mU{\mbox{$\mathbf{U}$}}

\newtheorem{definition}{Definition}

\newcommand{\beq}{\begin{equation}}
\newcommand{\eeq}{\end{equation}}
\def\BibTeX{{\rm B\kern-.05em{\sc i\kern-.025em b}\kern-.08em
    T\kern-.1667em\lower.7ex\hbox{E}\kern-.125emX}}
\begin{document}
\title{Cross-Laplacians Based Topological Signal Processing  over Cell MultiComplexes\\
{\footnotesize 
\thanks{This work was funded by the FIN-RIC Project %“Topological Signal Processing: Applications in bRain and Knowledge networks” 
TSP-ARK, financed by Universitas Mercatorum under grant n. 20-FIN/RIC.
}
}}

% \author{Stefania Sardellitti, Breno, Fernando and Lima  \\
% Universitas Mercatorum,  Engineering and Sciences Dept.,  Piazza Mattei 10, 00186, Rome, Italy\\ e-mail:  stefania.sardellitti@unimercatorum.it}

\author{\IEEEauthorblockN{Stefania Sardellitti$^1$, Breno C. Bispo$^2$, Fernando A. N. Santos$^3$, Juliano B. Lima$^2$}
\IEEEauthorblockA{{$^1$Engineering and Sciences Dept.}, {Universitas Mercatorum, Rome, Italy}
\\
{$^2$Dept. of Electronics and Systems}, {Federal University of Pernambuco, Recife, Brazil}\\
{$^3$Dutch Institute for Emergent Phenomena, KdVI}, {University of Amsterdam, Amsterdam, The Netherlands}
}
e-mail: stefania.sardellitti@unimercatorum.it, breno.bispo@ufpe.br, f.a.nobregasantos@uva.nl, juliano.lima@ufpe.br
}

\maketitle
%\and
%\and
%\and
%\and
%\and

%\maketitle

\begin{abstract} The study of the interactions among different types of interconnected systems in  complex networks  has attracted significant interest across many research fields. However, effective signal processing over layered networks requires topological descriptors of the intra- and cross-layers relationships  that are  able to disentangle the homologies of different domains,  at different scales, according to the specific learning task. In this paper, we  present Cell MultiComplex (CMC)  spaces,  which are novel topological domains for representing  multiple higher-order relationships among interconnected complexes. We introduce  cross-Laplacians matrices, which are algebraic descriptors of CMCs enabling the extraction of topological invariants at different scales, whether global  or local, inter-layer or intra-layer. Using the eigenvectors of these cross-Laplacians as signal bases, we develop topological signal processing tools for  CMC spaces. In this first study,
we  focus on the representation and filtering of noisy flows observed over cross-edges between different layers of CMCs  to identify cross-layer hubs, i.e., key nodes on one layer  controlling  the others. 
\end{abstract}


%maximum length 5 pages

\section{Introduction}
In recent years,  there has been a growing interest in the study of complex networks, as they model systems where  a set of entities interact  in different ways through relationships that often convey different meanings and scales \cite{boccaletti2006complex}.  Typically, complex systems are composed of multiple interconnected subsystems organized into distinct  layers of connectivity. 

Multilayer networks \cite{boccaletti2006complex}, \cite{de2013mathematical}, \cite{bianconi2021higher} have been extensively studied over the last few decades as  they provide a  natural and powerful framework for modeling heterogeneous systems. 
Unlike traditional single-layer networks, multilayer networks model multiple types of interactions within a  system, 
 by efficiently describing complex  phenomena. 
For instance, in neuroscience, multimodal brain connectomes can be modeled as multilayer networks \cite{10.1093/gigascience/gix004}, where different layers correspond to different modes of brain connectivity, potentially giving more nuances than single layer brain networks \cite{breedt2023multimodal}. In biological molecular networks \cite{liu2020robustness}, multiple biochemical interactions, as 
protein-gene-metabolite interactions, can be represented by multilayer networks. Similarly, in telecommunication and transportation networks \cite{CRAINIC20221} multilayers networks are efficient tools for analyzing  different levels of connectivities.  Most of these studies focused  on modeling  inter- and intra-layers relationships using graphs, which can only capture pairwise interactions between entities. However, in many complex systems, interactions typically involve groups of similar or heterogeneous entities,  leading to recent studies on higher-order multiplex networks \cite{krishnagopal2023topology} based on simplicial complexes. Simplicial complexes are topological spaces able to capture higher-order interactions between the elements of a set, while preserving the inclusion property.
Despite recent success in topological representations of complex systems via simplicial complexes, the current representation of these spaces rely on algebraic topological descriptors that fail to disentangle the  local intra- and inter-layers topological features.
In this regard, recently, the authors in \cite{moutuou2023} introduced an interesting representation  of simplicial multi-complex networks using the cross-Laplacians as algebraic topological descriptors. These matrices represent powerful algebraic tools for analyzing both global and local topological invariants of a space, i.e. properties that keep unchanged under homeomorphisms. These topological invariants are encoded by the so called cross-Betti vectors, i.e. a set of cross-Betti numbers able to capture different local topological invariants.

Our first novel contribution in this paper is extending the simplicial complex algebraic representation in \cite{moutuou2023} to cell complex spaces, which we  name \textit{Cell MultiComplexes (CMCs)}.   CMCs are powerful spaces capable of capturing multiple interactions  of any sparsity order among entities and that  can be efficiently represented through cross-Laplacians. %providing flexible topological structures able to grasp different topological invariants across different scales, global or local, depending on the data features we aim to reveal.
 By introducing  different boundaries maps,  cross-Laplacians enable the extraction of different kinds of topological invariants according to the scales we aim to explore: a global perspective, treating the entire complex as a flattened monolayer structure, or a local view, which  disentangles the homologies by investigating as the topology of one layer is related to 
 %controls
 the others.  
 
 Our second key contribution is the development of a signal processing framework on CMCs.
We  first introduce local Hodge decompositions of signals observed on the cells of a CMC,  enabling signal spectral representation. In this initial study, our learning-task focuses on  processing flows over the cross-edges connecting different layers in order to identify harmonic cross-hubs between layers. Then, we show how the homologies of  the $(0,0)$-cross Laplacians can effectively capture the number of harmonic cross-hubs between layers, i.e. key nodes controlling inter-layers connectivity. Using the eigenvectors of this cross-Laplacian as signal bases, we show how noisy flows across two different layers of a CMC can be efficiently  filtered  to recover the signal components that can be exploited in identifying cross-hubs.   
% In \cite{roddenberry2023signal} a framework for signal processing on product spaces of
 %simplicial and cellular complexes has been proposed.
 
\section{Cell multicomplexes}
\label{sec: Cell_multicomplexes}

In this section we introduce the fundamental notions defining cell multicomplexes. 
Building on the topological tools developed for simplicial complexes in \cite{moutuou2023}, our first novel contribution is to 
extend these representation methods to encompass more general topological structures, such as cell complexes. 
%In many applications, data resides on multiple interconnected domains or networks, where   the objetive of learning  is to uncover  global as well as local topological features. Thus,
 % representing data as a  monolayer topological structure may lead to a loss of key topological information. To address this, We here  propose a novel algebraic representation that is able to differentiate  between  the intra- and inter-layer homologies, following the approach introduced in \cite{moutuou2023}. This perspective allows to see each layer as exhibiting   different topological properties depending on how it is explored: whether we look at each layer from its point of view, through the lens of other layers, or as a part of a whole aggregate structure.\\
 We begin by recalling the notion of cell complexes 
 and then we introduce  cell multicomplexes topological spaces.\\
\textbf{Cell complexes.}
An  abstract cell complex (ACC) \cite{klette2000cell}
$\mathcal{C}=\{\mathcal{S},  \prec_b, \text{dim}\}$ is a set $\mathcal{S}$ of abstract elements $c$, named cells,
provided with a binary relation $\prec_b$, called the bounding (or incidence) relation, and with a dimension function, denoted by $\mbox{dim}(c)$,  assigning to  each $c \in \mathcal{S}$ a non-negative integer $[c]$.\\
A cell $c$ is called a $k$-cell if 
 $\text{dim}(c)=k$ where $k$ is the dimension (or order) of $c$. 
 We denote a cell of order $k$ as $c_k$. Therefore,  
 $0$-cells $c_0$ are named vertices and $1$-cells $c_1$ edges. We say that the $k$-cell $c_k$ lower bounds the $(k+1)$-cell $c_{k+1}$ 
 if $ c_k \prec_b c_{k+1}$
and $c_k$ is a face of $c_{k+1}$. % and $c_{k+1}$ is a co-face of $c_k$. 
%We also assume  every 1-cell of $\mathcal{C}$ 
% incident with two $0$-cells of $\mathcal{C}$  and distinct $1$-cells  not incident to the same pair of $0$-cells. 
 An ACC is of dimension $K$ or a $K$-dimensional ACC, if the dimensions of all its cells are less  than or equal to $K$. 
 Given a $k$-dimensional cell $c_k$,
 we define its boundary  as the set of all cells of dimension
 less than $k$  bounding $c_k$. \\
 %We  define   a closed-cell $\bar{c}_k$, as a cell which includes  its boundary, i.e. $\bar{c}_k= c_k \cup \partial c_k$. 
% An ACC equipped with neighboring relations among its cells is a
 %topological space \cite{barmak2011algebraic}.\\
 %As an example in Fig.\ref{fig:CC} we illustrate a cell-complex of order $3$, consisting of $12$ cells of order $2$ (triangles and squares) and $1$ cell of order $3$ (violet tetrahedron).
 %\begin{figure}[t]
%\centering
%\includegraphics[width=6.7cm,height=2.8cm]{CC.jpg}
%\caption{ An example of an ACC of order $3$.}
%\label{fig:CC}
%\end{figure}
\textbf{Cell Multicomplexes.}
Let us now introduce the concept of cell multicomplex space.
\begin{definition}
\textit{A Cell MultiComplex (CMC) $\mathcal{X}$ is a topological space composed by a finite collection  of interdependent abstract cell complexes, each  associated with a topological layer. The interdependence among these complexes involve higher-order 
inter-layer interactions modeled by cross-complexes.}    \end{definition}
 %For example, every layer can be associated with a different domain or with a different snapshot of the same domain, so that the cell multicomplex can be representative of the relationships among signals within or between different domains or networks.\\
 
The inter-layer higher-order interactions  are captured by cells of different orders  named \textit{cross-cells}.  The dimension of  a CMC is the  maximum order of its cross-cells. Cross-cells of order $1$, $2$ and $3$ are cross-edges, cross-polygons and cross-polyhedra, respectively.\\% However, we use these terms to denote abstract cells that are not necessarily embedded in a geometric space.\\
\begin{figure}[t]
\centering
\includegraphics[width=8.9cm,height=4.8cm]{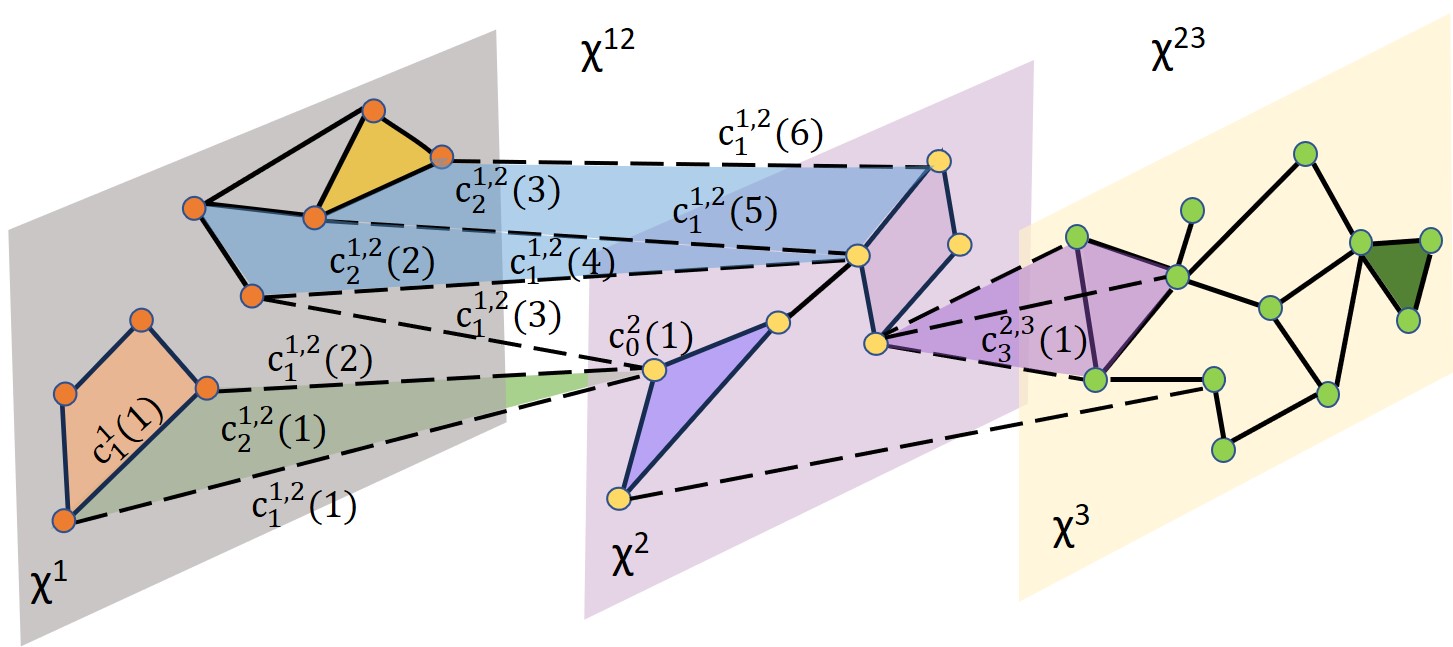}%{MC1n2.jpg}
\caption{Illustrative example of a CMC of order $3$.}
\label{fig:CMC_1}
\end{figure}
An illustrative example of a CMC  composed of $L=3$ layers is shown in  Fig. \ref{fig:CMC_1}.  It is composed of three intra-layer cell complexes $\mathcal{X}^{1}, \mathcal{X}^{2},\mathcal{X}^{3}$ interconnected by cross-edges (dashed lines). We can observe three cross-cells of order $2$ between layers $1$ and $2$, one  triangle and two squares,  and one cross-cell of order $3$, a  tetrahedron, between layer $2$ and $3$.
%Note that the cell complex in Fig. \ref{fig:CC} is indeed a monolayer perspective of the CMC in Fig. \ref{fig:CMC_1}. 
Note that  CMCs  are suitable spaces  to represent data observed over higher-order interconnected networks or over different domains each associated with a different layer. \\ %between different functionalities of the layers and their inter-connectivity. To address this, it is essential to capture also local topological features or invariants by introducing different algebraic representations tailored to the objective of the learning task.\\
Consider a network with layers indexed according to an  increasing order. For simplicity of notation and w.l.o.g., let us assume that cross-cells involve only two layers,   as illustrated in Fig. \ref{fig:CMC_1}. Hence, we denote by  $c_{k}^{\ell,m}(n)$  the $n$-th cross-cell of order  $k \geq 0$, interconnecting layers $\ell$ and $m$.
 According to this notation, the cells $c_{k}^{i}$ are intralayer cells, i.e. cells of order $k$ within layer $i$.  
 Given the cross-cell $c_{k}^{\ell,m}(n)$, we define its $\ell$-layer face and $m$-layer face  as the cells of order $0 \leq j<k$ that lower bound $c_{k}^{\ell,m}(n)$ and belong to the $\ell$-layer and $m$-layer, respectively. 
 %The faces of each cross-cell ${c_{k,n}}^{(1,2,\ldots,m)}$ are intra- and cross-cells of order $k-1$. 
  In the  example  in Fig. \ref{fig:CMC_1}, the 2-order cross-cell $c_2^{1,2}(1)$   is a cross-triangle, connecting layers $1$ and $2$,  with lower bounding $1$-order cells $c_{1}^{1}(1),c_{1}^{1,2}(1)$ and $c_{1}^{1,2}(2)$. The  face on layer $1$ of $c_2^{1,2}(1)$ is the edge $c_1^{1}(1)$, while the face on layer $2$ is the node $c_0^{2}(1)$. Note that cross-cells may have in general faces of different orders on each layer.\\  
%Similarly, the $2$-order cell $c_2^{1,2}(3)$ is a square cell with bounding edges $c_1^{1}(2),c_1^{1,2}(5),c_1^{1,2}(6),c_1^{2}(5)$ and $1$-order faces on layers $1$ and $2$, given by the cells $c_1^{1}(2)$ and $c_1^{2}(5)$.   
A  CMC $\mathcal{X}$  is   a collection of intra- and cross-layer complexes.  %$\mathcal{X}^{\mathcal{I}_i}$,  denoting with  $\mathcal{I}_{i}$ the set of indices of the layers that the crosscomplex  connects.
We denote by $\mathcal{X}^{\ell,m}$  the cross-complex composed of the  cross-cells inter-connecting layers $\ell$ and $m$.  We further define the cross-complex $\mathcal{X}_{k,n}^{\ell,m} \subseteq \mathcal{X}^{\ell,m}$ as the collection of cross-cells with faces of order $k$ in layer $\ell$ and with faces of order $n$ in layer $m$. Thus, we denote by $N_{k,n}^{\ell,m}$ the number of cross-cells in $\mathcal{X}_{k,n}^{\ell,m}$, i.e. $N_{k,n}^{\ell,m}=|\mathcal{X}_{k,n}^{\ell,m}|$. Additionally, we define  $\mathcal{X}_{k,-1}^{\ell,m}$ as    the   intra-layer cell complex of order $k$ within layer $\ell$, using the subscript $-1$ to indicate the absence of cells over layer $m$.

%To better illustrate the cross-complexes composing a given CMC,
Considering the example in Fig. \ref{fig:CMC_1}, %over each layer we have the intra-layer  $2$-order cell complexes  $\mathcal{X}^{1}$, $\mathcal{X}^{2}$ and $\mathcal{X}^{3}$ defined by  $\mathcal{X}^{i}=(\mathcal{V}_i, \mathcal{E}_i,\mathcal{C}_i)$ with $|\mathcal{V}_i|=N_i$, $|\mathcal{E}_i|=E_i$, $|\mathcal{C}_i|=C_i$, respectively, the number of 0-order (vertices), 1-order (edges) and  2-order (polygonal) cells in the $i$-th layer. 
the cross-complex $\mathcal{X}^{1,2}$ between layer $1$ and $2$, is given by
 $\mathcal{X}^{1,2}=\{\mathcal{X}^{1,2}_{0,0},\mathcal{X}^{1,2}_{1,0},\mathcal{X}^{1,2}_{1,1}\}$ with  $\mathcal{X}^{1,2}_{0,0}= \{c_{1}^{1,2}(i)\}_{i=1}^{6}$, $\mathcal{X}^{1,2}_{1,0}=\{c_2^{1,2}(1),c_2^{1,2}(2)\}$,  and 
 $\mathcal{X}^{1,2}_{1,1}= \{c_2^{1,2}(3)\}$. Note that the complex $\mathcal{X}_{0,0}^{1,2}$ is a cross-graph, while $\mathcal{X}^{1,2}_{1,0}$ and $\mathcal{X}^{1,2}_{1,1}$ are cross-complexes of order $2$.

 The orientation of the cross-cells is an ordering choice over its lower bounding cells (see \cite{grady2010}, \cite{sardellitti2024topological}).
%Every simplex can only  have two orientations, depending on the permutations of its elements. two orientations are equivalent if each of them can be recovered from the other through an even number of transpositions,  where a transposition is a  permutation of two elements \cite{munkres2000topology}.\\
%To define the orientation of $k$-crosscells, we may apply first a simplicial decomposition \cite{grady2010}, which consists in partitioning the cell into a set of internal $k$-simplices, in a way such that: i) two simplices share exactly one $(k-1)$-simplicial boundary element, which is not the boundary of any other $k$-cell in the complex; and ii) two $k$-simplices induce an opposite orientation on the shared $(k-1)$-boundary. Therefore, by orienting  each internal simplex, the orientation propagates on the entire cell.\\ 
We use the notation $c_{k-1}^{\ell,m}(i)  \sim c_{k}^{\ell,m}(j)$ to indicate that the orientation of $c_{k-1}^{\ell,m}(i)$ is coherent with that of  $c_{k}^{\ell,m}(j)$ and $c_{k-1}^{\ell,m}(i) \nsim c_{k}^{\ell,m}(j)$ to indicate opposite orientations. %An oriented $k$-order crosscell $c_{k}^{\ell,m}$ of a CMC may be represented through its bounding cells  with two consecutive $(k-1)$-cross cells, sharing a common $(k-2)$-crosscell boundary. %For example, the oriented crosscell $c_2^{1,2}(2)$ in Fig. \ref{fig:CMC_12} can be denoted as $c_2^{1,2}(2)=[c_1^{1}(2),c_1^{1,2}(6), c_{1}^{2}(5),c_{1}^{1,2}(5)]$
%Given an orientation, there are two ways in which two crosscells can be considered to be adjacent: lower and upper adjacent.
Two $k$-order cells are lower adjacent if they share a common face of order $k-1$ and upper adjacent if they are both faces of a cell of order $k+1$. %\vspace{-0.3cm}

 \section{Algebraic representation of CMCs: from global to local invariants}
\label{sec: Alg_rep}
In many applications,  from signal processing to machine learning,  data resides on different interconnected  networks and   the learning task  is to uncover  global as well as local topological features. Therefore,
 we need a   framework for processing  signals defined over cell multicomplexes capable of disentangling the homologies of individual layers and uncovers how one layer influences and controls one another.   Depending on the learning task, we can adopt two main approaches for the analysis of  signals over a cell multicomplex. In the first common approach,  the entire structure is treated as a monolayer topological domain, so that the Hodge-Laplacian matrix introduced for cell complexes  can be used for the representation and processing of signals \cite{sardellitti2024topological}. In the second, novel approach,  we leverage cross-Laplacian matrices for  signal representations able to capture intra- and inter-layer homologies and uncover local topological invariants within the complex.
 
 \subsection{Cell multicomplex as a monolayer cell-complex} 
 One of the most common approaches in the study of multilayer networks is to represent them as a single monolayer structure. 
 %This allows  to capture the topological invariants of the overall structure by seeing the network as an unique entity.  
 Then, the resulting flattened cell complex  can be algebraically represented  by using the Hodge-Laplacian matrix \cite{sardellitti2024topological}.
Let us assume that a $2$-order multicomplex $\mathcal{X}$ is composed of $L$ interconnected  layers. 
%Denote by  $\mathcal{G}^{l}=(\mathcal{N}_l,\mathcal{E}_l)$ the (intra-)layer graph 
%where  $\mathcal{N}_l$ and $\mathcal{E}_l$ are, respectively, the set of $N_l$ nodes on layer $l$
%and the set of $E_l$ edges on layer $l$.
% We represent by $\mathcal{X}^{l,m}_{0,0}=\mathcal{E}_{l,m}$ the set of $1$-cells (or cross-edges) connecting layer $l$ and $m$ with $|\mathcal{E}_{l,m}|=E_{l,m}$.  Then, the CMC $\mathcal{X}$ is considered as a single cell complex whose underlying graph is defined as $\mathcal{G}=(\mathcal{V},\mathcal{E})$, with a total number of nodes  $|\mathcal{V}|=N$ and a number of  edges $|\mathcal{E}|=E$. Specifically, it holds  $N=\sum_{l=1}^{L} N_{l}$ and 
 %$E=\sum_{l=1}^{L} E_{l}+\sum_{l=1}^{L}\sum_{m=1, m>l}^{L} E_{l,m}$. Furthermore, the number $C$ of overall $2$-order cells is given by the sum of the intra- and inter-layers $2$-cells.
The incidence matrix $\mB_k$, describing which $k$-cells are upper adjacent to which $(k-1)$-cells  is defined as $B_k(i,j)=1$ (or $B_k(i,j)=-1$) if $c_{k-1}(i) \prec_b c_{k}(j)$ and  $c_{k-1}(i) \sim c_{k}(j)$ (or $c_{k-1}(i) \not\sim c_{k}(j)$), while $B_k(i,j)=0$ if $c_{k-1}(i) \not\prec_b c_{k}(j)$.
Therefore, we can represent the cell multicomplex $\mathcal{X}$ using the graph Laplacian matrix $\mL_0=\mB_1\mB_1^T$ and the first-order Hodge Laplacian matrix $\mL_1=\mB_1^T\mB_1+\mB_2\mB_2^T$ \cite{sardellitti2024topological}.

%\beq
%\begin{array}{lll}
%\mL_0=\mB_1\mB_1^T, \medskip\\
%\mL_1=\mB_1^T\mB_1+\mB_2\mB_2^T
%\end{array}
%\eeq
%where $\mL_0$ is the graph Laplacian matrix, while $\mL_1$ is the first-order Laplacian matrix \cite{sardellitti2024topological}.\\
This representation of the CMC provides global invariants of the topological spaces described by the Betti numbers. Specifically,  $\beta_{0}=\text{dim}(\text{ker}(\mL_0))$  represents the number of connected components of the multilayer graph, while   $\beta_{1}=\text{dim}(\text{ker}(\mL_1))$ corresponds to  the number of holes in the entire complex, i.e. the number of empty $2$-cells within the complex.
%It has to be remarked that the 0-order Laplacian of the CM $\mathcal{X}$ is given by:
%\beq
%\mB_1=\left[ \begin{array}{lll}
%\mB_{0,-1}^{(1),2} & \mB_{0,-1}^{1,(2)} & \mathbf{0}\\
%\mathbf{0} & \mB_{-1,0}^{(1),2} & \mB_{-1,0}^{1,(2)} 
%\end{array}\right]
%\eeq
%and
%\beq
%\mB_1 \mB_1^T=\left[ \begin{array}{lll}
%\mB_{0,-1}^{(1),2}(\mB_{0,-1}^{(1),2})^T + \mB_{0,-1}^{1,(2)}(\mB_{0,-1}^{1,(2)})^T & \mB_{0,-1}^{1,(2)} ( \mB_{-1,0}^{(1),2})^T\\
 %\mB_{-1,0}^{(1),2} (\mB_{0,-1}^{1,(2)} )^T & \mB_{-1,0}^{(1),2}(\mB_{-1,0}^{(1),2})^T+\mB_{-1,0}^{1,(2)} (\mB_{-1,0}^{1,(2))} 
%\end{array}\right]
%\eeq
%the Betti number is $\beta_0=1$ since it is a  connected overall graph.
%In many other work (Battiston et al. 2014), a monoplex network is constructed by aggregating data from the different layers of a multilayer network, the classical definition of node degree is then applied to the resulting monoplex network. However, network aggregation leads to a loss of information. In Some other work, the distinction of the layers is maintained and the degree of node is represented by a vector.

 \subsection{Cross-Laplacians to capture cross-invariants}
%Albeit, representing the cell complex as a single multi-complex can be useful in contexts where we are interested to global invariants of the space, this representation fails to capture specific features from data. We need a topological representation of data that is task-oriented, i.e. 
%   able to distinguish the interdependence between different domains and see how the topology of a layer affects and controls the topology of other layers.   
In this section, we introduce the notion of cross-Laplacians matrices presented in \cite{moutuou2023} by extending it to cell multi-complexes.
 For simplicity of notation,  let us assume that cross-cells involve only pairs of layers. \\%This allows simplifying the notational presentation of our algebraic tools albeit the extension to the case where inter-cells  connect multiply layers is straightforward. \\
\textbf{Cross-boundaries maps.}
  First we introduce the boundaries maps of cross-cells in the perspective of  a given layer, i.e. the boundaries maps of cross-cells only with respect to faces  belonging to a given layer  and keeping fixed all the remaining faces.  
  Let us consider the two layers $\ell,m$ and denote by $C^{k,n}$ the real vector space generated by  all oriented $q$-order cross-cells $c_q^{\ell,m}$, with faces of order $k$ on layer $\ell$ and faces of order $n$  on layer $m$.
  To simplify our notation, we omit the dependence of the cell's order $q$  on the orders $(k,n)$ of the cells on  layers $\ell$ and $m$, respectively. As an example, in a $2$-order CMC, if $(k,n)=(0,0)$, we obtain $q=1$, corresponding to cross-edges, while for $(k,n)=(1,0),(0,1),(1,1)$, we have  $2$-order cross-cells.  Hence, given the cross-complex $\mathcal{X}_{k,n}^{\ell,m}$  we can define two distinct cross-boundaries operators for each cross-cell $c_{q}^{\ell,m}\in \mathcal{X}_{k,n}^{\ell,m}$. The first operator $\mB_{k,n}^{(\ell),m}$ is a boundary map defined with respect to the crossfaces on layer $\ell$, while the second operator, denoted as $\mB_{k,n}^{\ell,(m)}$, is a boundary map with respect to the crossfaces on layer $m$.
  Specifically, $\mB_{k,n}^{(l),m}: C^{k,n} \rightarrow C^{k-1,n}$ is  the boundary map with respect to the cells of order $k$ on layer $\ell$ as view from layer $m$. Thus, $\mB_{k,n}^{(\ell),m}$ is an incidence matrix of dimension $N_{k-1,n}^{\ell,m} \times N_{k,n}^{\ell,m}$ with entries defined as 
    \beq \label{eq:B_kn_ell}
  B_{k,n}^{(l),m}(i,j)\!=\!\!\left\{\!\!\!\begin{array}{rll}
  0, & \! \text{if} \; c_{q-1}^{\ell,m}(i) \not\prec_b c_{q}^{\ell,m}(j) \medskip\\
  1,& \!\text{if} \; c_{q-1}^{\ell,m}(i) \prec_b c_{q}^{\ell,m}(j) ,  c_{q-1}^{\ell,m}(i) \sim c_{q}^{\ell,m}(j)\medskip\\
  -1,& \!\text{if} \; c_{q-1}^{\ell,m}(i) \prec_b c_{q}^{\ell,m}(j),   c_{q-1}^{\ell,m}(i) \nsim c_{q}^{\ell,m}(j)\\
  \end{array}\right. 
  \eeq
  where $c_{q-1}^{\ell,m}(i)\in \mathcal{X}_{k-1,n}^{\ell,m}$ and $c_{q}^{\ell,m}(j)\in \mathcal{X}_{k,n}^{\ell,m}$ , $\forall i,j$. 
   Similarly, the  matrices $\mB_{k,n}^{\ell,(m)}: C^{k,n} \rightarrow C^{k,n-1}$ of dimension 
   $N_{k,n-1}^{\ell,m} \times N_{k,n}^{\ell,m}$ are boundaries with respect to crossfaces in layer $m$ with entries $B_{k,n}^{l,(m)}(i,j)$ defined as in  (\ref{eq:B_kn_ell}), except that $c_{q-1}^{\ell,m}(i)\in \mathcal{X}_{k,n-1}^{\ell,m}$ and  $c_{q}^{\ell,m}(j)\in \mathcal{X}_{k,n}^{\ell,m}$, $\forall i,j$. 
  %  \beq \label{eq:B_kn_m}
 % B_{k,n}^{l,(m)}(i,j)\!=\!\!\left\{\!\!\!\begin{array}{rll}
 % 0, & \! \text{if} \; c_{q-1}^{\ell,m}(i) \not\prec_b c_{q}^{\ell,m}(j) \\
 % 1,& \! \text{if} \; c_{q-1}^{\ell,m}(i) \prec_b c_{q}^{\ell,m}(j), c_{q-1}^{\ell,m}(i) \sim c_{q}^{\ell,m}(j)\\
 % -1,& \! \text{if} \; c_{q-1}^{\ell,m}(i) \prec_b c_{q}^{\ell,m}(j),  c_{q-1}^{\ell,m}(i) \nsim c_{q}^{\ell,m}(j)\\
%  \end{array}\right. 
 % \eeq
  It can be proved (we omit here the proof for lack of space) that
  \beq \label{eq:orth}  \mB_{k,n}^{(\ell),m}\mB_{k+1,n}^{(\ell),m}=\mathbf{0} \quad \text{and} \quad \mB_{k,n}^{\ell,(m)}\mB_{k,n+1}^{\ell,(m)}=\mathbf{0}. \eeq  
As an example, let us consider the cross-complex $\mathcal{X}_{1,0}^{1,2}=\{c_2^{1,2}(1),c_2^{1,2}(2)\}$ in Fig. \ref{fig:CMC_1}. This complex  consists of two  cross-cells of order $q=2$,  i.e., $c_2^{1,2}(1)$  and $c_2^{1,2}(2)$. These two cells have faces of order $1$ (edge or paths) on layer $1$ and  one face of order $0$ (vertex) on layer $2$. The bounding cells of $c_{2}^{1,2}(1)$   are: with respect to cells on layer $1$, the two cross-edges $c_{1}^{1,2}(1)$ and $c_{1}^{1,2}(2)$, while with respect to cells on layer $2$ the bounding cell is $c_1^{1}(1)$.  \\
%The number of $2$-order cross-cells  is $N_{1,0}^{1,2}=\mid \mathcal{X}_{1,0}^{1,2}\mid=2$ and the number of cross-edges is $N_{0,0}^{1,2}=\mid \mathcal{X}_{0,0}^{1,2}\mid=6$. %Therefore, the   matrix $\mB_{1,0}^{(1),2} \in \mathbb{R}^{6\times 2}$  will be derived as (\ref{eq:B_kn_ell}).\\% as 
%\beq \label{eq:ex1}
%\mB_{1,0}^{(1),2}=\left[\begin{array}{rrrrrrr}-1 & 1 & 0&0&0&0\\
%0 & 0& 0& -1& 1& 0\end{array}\right]^T.
%\eeq
%Similarly, the matrix $\mB_{1,0}^{1,(2)} \in \mathbb{R}^{10 \times 2}$, with $N_{1,-1}^{1,2}=10$, will be 
%\beq \label{eq:ex2}
%\mB_{1,0}^{1,(2)}=\left[\begin{array}{rrrrrrrrrrr}1 & 0 & 0&0&0&0&0&0&0&0\\
%0 & 0& 0& 0& 1& -1& 0&0&0&0\end{array}\right]^T.
%\eeq
%In summary, from (\ref{eq:ex1}), we can notice that the bounds of a cross-cell, with faces $k=1,n=0$, with respect to layer $1$ are the two cross-edges, while with respect to layer $2$, are the intra-edges on layer $1$. \\
\textbf{Cross-Laplacian matrices.}  
Given the two layers $\ell,m$, we  introduce  the  $(k,n)$-cross-Laplacian matrices from layer $\ell$  as
\beq \label{eq:Lknl_l}
\mL_{k,n}^{(\ell),m}=(\mB_{k,n}^{(\ell),m})^{T}\mB_{k,n}^{(\ell),m}+\mB_{k+1,n}^{(\ell),m} (\mB_{k+1,n}^{(\ell),m})^{T}
\eeq
%\beq \label{eq:Lknl_l}
%\mL_{k,n}^{(\ell),m}=\underbrace{{(\mB_{k,n}^{(\ell),m})^{T}\mB_{k,n}^{(\ell),m}}}_{\text{Lower Laplacian from layer $\ell$}} +\underbrace{\mB_{k+1,n}^{(\ell),m} (\mB_{k+1,n}^{(\ell),m})^{T}}_{\text{Upper Laplacian from layer $\ell$}}  
%\eeq
where the first and second terms encode the lower and upper adjacencies, respectively. 
Similarly the  $(k,n)$-cross-Laplacian matrices from layer $m$ are
\beq \label{eq:Lknl_m}
\mL_{k,n}^{\ell,(m)}=(\mB_{k,n}^{\ell,(m)})^{T}\mB_{k,n}^{\ell,(m)} +\mB_{k,n+1}^{\ell,(m)}(\mB_{k,n+1}^{\ell,(m)})^T.
\eeq
%\beq \label{eq:Lknl_m}
%\mL_{k,n}^{\ell,(m)}=\underbrace{(\mB_{k,n}^{\ell,(m)})^{T}\mB_{k,n}^{\ell,(m)}}_{\text{Lower Laplacian from layer $m$}}  +\underbrace{\mB_{k,n+1}^{\ell,(m)}(\mB_{k,n+1}^{\ell,(m)})^T}_{\text{Upper Laplacian from layer $m$}}.
%\eeq
These Laplacians matrices are symmetric and semidefinite positive. 
It can be observed that the intra $\ell$-layer Hodge Laplacian of order $k$  can be derived from (\ref{eq:Lknl_l}) by setting $n=-1$. %, as
%\beq \label{eq:Lk-1}
%\mL_{k,-1}^{(\ell),m}= (\mB_{k,-1}^{(\ell),m})^{T}\mB_{k,-1}^{(\ell),m}+\mB_{k+1,-1}^{(\ell),m}(\mB_{k+1,-1}^{(\ell),m})^T. 
%\eeq
%Similarly, the layer $m$ Hodge Laplacian of order $k$ is obtained from (\ref{eq:Lknl_m}) as
%\beq \label{eq:L-1k}
%\mL_{-1,k}^{l,(m)}=(\mB_{-1,k}^{l,(m)})^{T} \mB_{-1,k}^{l,(m)} + \mB_{-1,k+1}^{l, (m)} (\mB_{-1,k+1}^{l,(m)})^{T}.
%\eeq
Additionally, note that it holds 
$\mB_{k,-1}^{\ell,(m)}=\mathbf{0}, \;  \; \mB_{-1,n}^{(\ell),m}=\mathbf{0}, \; \forall k,n$.
Furthermore, from (\ref{eq:orth}), it can be proved, following similar considerations as in \cite{Lim,moutuou2023}, that the space $\mathbb{R}^{N_{k,n}}$ admits different Hodge decompositions according to the layer from which the boundary is calculated. Specifically, it holds
\beq \label{eq:R_Nkn_ell}
\mathbb{R}^{N_{k,n} } \equiv \text{img}(\mB_{k,n}^{(\ell),m \,T}) \oplus \text{ker}(\mL_{k,n}^{(\ell),m}) \oplus  \text{img}(\mB_{k+1,n}^{(\ell),m}),
\eeq
\beq \label{eq:R_Nkn_m}
\mathbb{R}^{N_{k,n} } \equiv \text{img}(\mB_{k,n}^{\ell,(m) \,T}) \oplus \text{ker}(\mL_{k,n}^{\ell,(m)}) \oplus  \text{img}(\mB_{k,n+1}^{\ell,(m)}).
\eeq
The orthogonality conditions in (\ref{eq:orth}) allow to define the  $(k,n)$-cross-homology groups of $\mathcal{X}$ \cite{Lim,moutuou2023}, as $\text{H}_{k,n}^{(\ell)}\cong \text{ker}(\mL_{k,n}^{(\ell),m})$ and $\text{H}_{k,n}^{(m)}\cong \text{ker}(\mL_{k,n}^{\ell,(m)})$. The cross-homology groups are determined by their dimensions, named the  $(k,n)$-cross-Betti numbers \cite{moutuou2023}, $\beta_{k,n}^{(\ell)}= \text{dim} (\text{ker}(\mL_{k,n}^{(\ell),m}))$ and $\beta_{k,n}^{(m)}= \text{dim} (\text{ker}(\mL_{k,n}^{\ell,(m)}))$. Then, we can define the $(k,n)$-cross-Betti vector of $\mathcal{X}_{k,n}^{\ell,m}$ as the vector 
$\boldsymbol{\beta}_{k,n}^{\ell,m}=[\beta_{k,n}^{(\ell)},\beta_{k,n}^{(m)}]$.
These numbers, as we will see for a $2$-order CMC, are able to capture the homologies of the intra- and cross-layer cell complexes.

\section{Second-order Cell MultiComplexes}

Considering a $2$-order CMC, we can build different cross-Laplacians according to the topological invariants we aim to detect.  In this first study we %\subsection{The $(0,0)$-cross-Laplacians} Let us first 
focus on the  
$(0,0)$-cross-Laplacians. %$\mL_{0,0}^{(\ell),m}$. 
Using (\ref{eq:Lknl_l}) the Laplacian $\mL_{0,0}^{(\ell),m}$ is an %$N_{0,0}\times N_{0,0}$-symemtric matrix with $|N_{0,0}|=|\mathcal{X}_{0,0}^{\ell,m}|$the
 $N_{0,0}^{l,m}\times N_{0,0}^{l,m}$ symmetric matrix indexed on the cross-edges $c_1^{l, m} \in \mathcal{X}^{l,m}_{0,0}$ expressed as 
\beq \label{eq:L_00_l_m}
\mL_{0,0}^{(l),m}= (\mB_{0,0}^{(l),m})^{T} \mB_{0,0}^{(l),m}+\mB_{1,0}^{(l),m}(\mB_{1,0}^{(l),m})^{T}. 
\eeq
Using 
 (\ref{eq:B_kn_ell}), we  get the $N_{-1,0}^{\ell,m}\times N_{0,0}^{l,m}$ incidence matrix 
  \beq \label{B_00l}
  B_{0,0}^{(l),m}(i,j)=\left\{\begin{array}{rll}
  0, &  \text{if} \; c_{0}^{m}(i) \not\prec_b c_{1}^{l,m}(j) \\
  1,& \text{if} \; c_{0}^{m}(i) \prec_b c_{1}^{l,m}(j),\; c_{0}^{m}(i) \sim  c_{1}^{l,m}(j)\\
  -1,& \text{if} \; c_{0}^{m}(i) \prec_b c_{1}^{l,m}(j), \;    c_{0}^{m}(i) \nsim  c_{1}^{l,m}(j)\\
  \end{array}\right. 
  \eeq
  with $c_0^{m}(i) \in \mathcal{X}_{-1,0}^{\ell,m}$ and $c_1^{\ell,m}(j) \in \mathcal{X}_{0,0}^{\ell,m}$. 
Then, the entry $(i,j)$ of the lower Laplacian  $(\mB_{0,0}^{(l),m})^T\mB_{0,0}^{(l),m}$  is equal to  $1$ if $c_{1}^{l,m}(i)$ is lower adjacent to $c_{1}^{l,m}(j)$ on layer $m$, i.e.  $c_{1}^{l,m}(i)$ and $c_{1}^{l,m}(j)$ have a common vertex on layer $m$. The incidence matrix $\mB_{1,0}^{(l),m}: C^{1,0} \rightarrow C^{0,0}$, in the second term of (\ref{eq:L_00_l_m}),
is a $N_{0,0}^{l,m}\times N_{1,0}^{\ell,m}$  matrix with $N_{1,0}^{\ell,m}$ being the number of cross-cells of order $2$ between layer $\ell,m$ having edges over layer $\ell$ and one vertex over layer $m$. 
%For example, if we have a triangle $c_{2}^{l,m}(j)=\{c_{1}^{l}(i),c_{1}^{l,m}(k),c_{1}^{l,m}(n)\}$  with one edge belonging to layer $l$,
 Then, we get from (\ref{eq:B_kn_ell}):
\beq \label{B_10l}
  B_{1,0}^{(l),m}(i,j)\!=\!\!\left\{\begin{array}{rll}
  0, & \text{if} \; c_{1}^{l,m}(i) \not\prec_b c_{2}^{l,m}(j) \\
 % 0, & \text{if} \; c_{1}^{l}(i) \not\prec_b c_{2}^{l,m}(j) \\
  1,& \text{if} \; c_{1}^{l,m}(i) \prec_b c_{2}^{l,m}(j),  c_{1}^{l,m}(i) \sim  c_{2}^{l,m}(j)\\
  -1,& \text{if} \; c_{1}^{l,m}(i) \prec_b c_{2}^{l,m}(j),    c_{1}^{l,m}(i) \nsim  c_{2}^{l,m}(j)\\
  \end{array}\right. 
  \eeq
for $c_{1}^{l,m}(i) \in  \mathcal{X}_{0,0}^{\ell,m}$ and $c_{2}^{l,m}(j) \in  \mathcal{X}_{1,0}^{\ell,m}$. 
   The upper Laplacian
$\mB_{1,0}^{(l),m}(\mB_{1,0}^{(l),m})^{T}$ identifies the upper adjacencies of the cross-edges $c^{l,m}_1$ as boundaries of $2$-order cells with edges on layer $\ell$ and one vertex on layer $m$. Similar derivations can be followed to obtain the cross-Laplacian $\mL_{0,0}^{\ell,(m)}$.\\
\textbf{The Cross-Betti  vector $\boldsymbol{\beta}_{0,0}$.}
To describe the topological invariants encoded by the cross-Betti vector  $\boldsymbol{\beta}_{0,0}$, we need to introduce the concept of  cones \cite{moutuou2023}. Cones are   the shortest paths of length two between nodes within one layer, passing through a node on the other layer and not belonging to the cross-boundary of $2$-order cross-cells.
The cones are called closed if they form a cycle. For example, a cycle can have one vertex  on layer $m$ and the remaining  vertices  on layer $\ell$.
A cone can also be open,  meaning that a vertex on one layer connects   clusters on the other layer that are unconnected.
Then, the cross-Betti number $\beta_{0,0}^{(l)}=\text{ker}(\mL_{0,0}^{(l),m})$ is equal to the number of  cones (closed and open), with one vertex on layer $m$, that are not boundaries of $2$-order cross-cells.
The vertices of the cones  on layer $m$ are named harmonic cross-hubs. Similarly, $\beta_{0,0}^{(m)}=\text{ker}(\mL_{0,0}^{l,(m)})$
identify the number of cones with one vertex on layer $\ell$.
%Similar considerations hold for the cross-Betti number $\beta_{0,0}^{(m)}=\text{ker}(\mL_{0,0}^{l,(m)})$.

%\subsection{The $(1,-1)$ cross-Laplacians}
%Let us now consider the $\mL_{1,-1}^{(\ell),m}$ cross-Laplacian matrix. From  (\ref{eq:Lknl_l}), we get
%\beq \label{eq:Lknl_1_1}
%\mL_{1,-1}^{(l),m}=(\mB_{1,-1}^{(l),m})^{T}\mB_{1,-1}^{(l),m} +\mB_{2,-1}^{(l),m} (\mB_{2,-1}^{(l),m})^{T},
%\eeq
%that is the classical first-order Hodge Laplacian of the $\ell$  intra-layer  complex.
%On the other hand, using (\ref{eq:Lknl_m}), we get
%\beq \label{eq:Lknl_m_1}
%\mL_{1,-1}^{l,(m)}=\mB_{1,0}^{l,(m)}(\mB_{1,0}^{l,(m)})^T,
%\eeq
%where $\mL_{1,-1}^{l,(m)}$  is a $N_{1,-1}\times N_{1,-1}$  diagonal matrix with entries the upper degree of the edges on layer $\ell$ bounding cross-cells. Therefore, $\beta_{1,-1}^{\ell, (m)}$ counts the number of edges in layer $\ell$ that are not $\ell$ faces of $2$-order cross cells, while $\beta_{1,-1}^{(\ell), m}$ counts the number of void $2$-order cells in the complex $\mathcal{X}^{\ell}$.

\section{Signal Processing over CMCs} 
Algebraic representations of CMCs  derived from cross-Laplacians offer suitable bases for the  processing of signals defined over CMCs.  
Let us consider a $2$-order CMC $\mathcal{X}=(\mathcal{V},\mathcal{E},\mathcal{C})$, with $\mid \mathcal{V}\mid=N$,  $\mid \mathcal{E}\mid=E$ and  $\mid \mathcal{C}\mid=C$ the dimension of the node, edges and $2$-cells sets, respectively. We can define signals over the set of nodes, edges and $2$-cells as $\bs_0: \mathcal{V}\rightarrow \mathbb{R}^{N}$, $\bs_1: \mathcal{E}\rightarrow \mathbb{R}^{E}$ and $\bs_2: \mathcal{C}\rightarrow \mathbb{R}^{C}$,  respectively. 
Using  the Hodge decompositions in (\ref{eq:R_Nkn_ell}), (\ref{eq:R_Nkn_m}),
we can split these signals  in different components belonging to orthogonal subspaces  and capturing distinct space invariants.

\begin{figure}[!t]
\centering
\includegraphics[width=8.9cm,height=4.8cm]{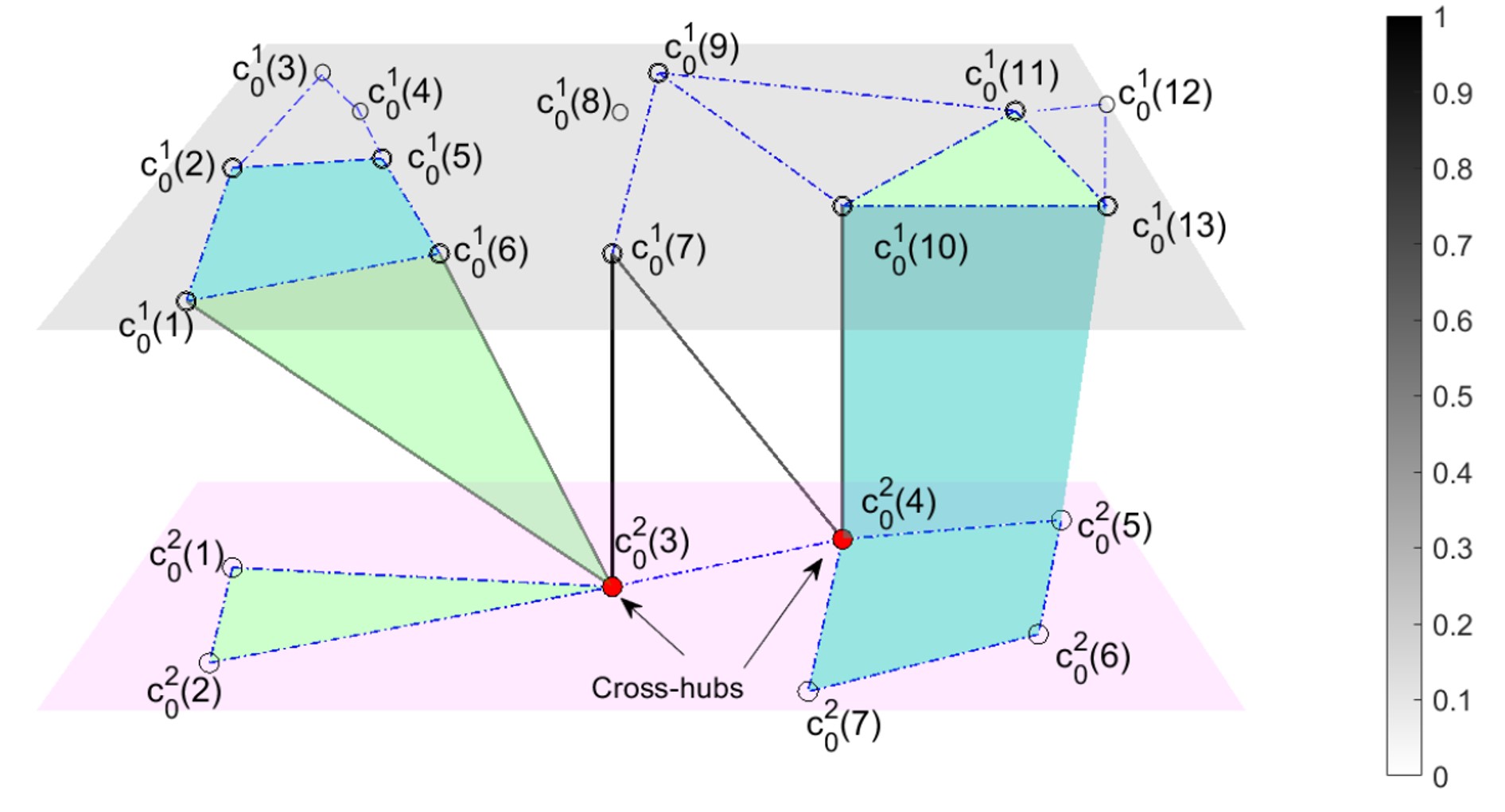}%{MC1n2.jpg}
\caption{Recovered harmonic cross-edge signals.}
\label{fig:CMC_2}
\end{figure}

%Then, we can define distinct decomposition of the  cross-edge signals $\bs_{1}^{\ell,m}$ according to the topological invariants that we aim to capture.
Let us focus on the  $(0,0)$-cross Laplacian in (\ref{eq:L_00_l_m}).  Then, it can be proved using (\ref{eq:R_Nkn_ell}) (we omit here the proof for lack of space), that the cross-edges signal $\bs_{1}^{\ell,m}$, belonging to the space $\mathbb{R}^{N_{0,0}^{\ell,m}}$, can be decomposed as
\beq \bs_1^{\ell,m} = \mB_{0,0}^{(\ell),m \, T} \bs_0^{m}+ \mB_{1,0}^{(\ell),m} \bs_2^{\ell,m}+\bs_{1,H}^{\ell,m},
\eeq
where the node signal $\bs_0^{m}\in \mathbb{R}^{N_{-1,1}^{\ell,m}}$ is observed over the nodes within layer $m$ and $\bs_2^{\ell, m}\in \mathbb{R}^{N_{1,0}^{\ell,m}}$ is a $2$-order signal observed over filled cones $(1,0)$ between layers $\ell,m$, i.e. cones with one vertex  on layer $m$. The first term  $\mB_{0,0}^{(\ell),m \, T} \bs_0^{m}$ is a flow on the cross-edges with zero-circulation along the cross-edges of filled cones $(1,0)$. The second flow $\mB_{1,0}^{(\ell),m} \bs_2^{\ell,m}$ has zero-sum on the vertices over layer $m$.  Finally, the harmonic edge signal  $\bs_{1,H}^{\ell,m}$ belong to the subspace spanned by  $\text{ker}(\mL_{0,0}^{\ell,m})$, whose dimension is the number of  (empty) cones between the two layers. 
Note that for the flows between layers $\ell$ and $m$, we can define 
%\beq \nonumber \vspace{-0.2cm}
%\text{div}^{(\ell),m}(\bs_{1}^{\ell,m})= \mB_{0,0}^{(\ell),m} \bs_1^{\ell,m}, \; \text{curl}^{(\ell),m}(\bs_{1}^{\ell,m})= \mB_{1,0}^{(\ell),m\, T} \bs_1^{\ell,m}\!. 
%\eeq
the term $\text{div}^{(\ell),m}(\bs_{1}^{\ell,m})=\mB_{0,0}^{(\ell),m} \bs_1^{\ell,m}$ that is a node signal measuring the conservation of the cross-flows  over the nodes of layer $m$, while $\text{curl}^{(\ell),m}(\bs_{1}^{\ell,m})= \mB_{1,0}^{(\ell),m\, T} \bs_1^{\ell,m}$ is a measure of the flow conservation along cross-edges bounding filled cone. \\
Extending the cell complex spectral theory \cite{sardellitti2024topological} to CMC  and given the eigendecomposition $\mL_{0,0}^{(\ell),m}=\mU_{0,0}^{(\ell),m} \boldsymbol{\Lambda}_{0,0}^{(\ell),m}\mU_{0,0}^{(\ell),m\, T}$, we can define the CMC Fourier Transform as the projection of a cross-edge signal $\bs_{1}^{\ell,m}$ onto the space spanned by the eigenvectors of $\mL_{0,0}^{(\ell),m}$, i.e.   $\hat{\bs}_{1}^{\ell,m}:= \mU_{0,0}^{(\ell),m\, T}\bs_{1}^{\ell,m} $. Hence, the cross-edge signal can be represented as
$
\bs_{1}^{\ell,m}:= \mU_{0,0}^{(\ell),m}\hat{\bs}_{1}^{\ell,m}.
$
Then,  %extending the TSP framework for cell complexes \cite{sardellitti2024topological} to CMCs,
we design optimal signal estimators from observed noisy cross-signals
$\by_1^{\ell,m}=\bs_1^{\ell,m}+\bn_1$,
where $\bn_1$ is additive noise. The optimal node, $2$-cells and harmonic   signals, can be derived as the solutions of the following problem
\beq \nonumber
\begin{array}{lllll}
 \!\!\underset{ \underset{{\bs}_{1,H}^{\ell,m} \in \mathbb{R}^{N_{0,0}}}{\bs_{0}^{m}\in \mathbb{R}^{N_m},{\bs}_{2}^{\ell,m} \in \mathbb{R}^{N_{1,0}}}}{\min} \!\! \!\!\!\!\parallel  \mB_{0,0}^{(l),m \, T} \!\bs_0^{m}\!\!+ \mB_{1,0}^{(l),m} \bs_2^{\ell,m}\!\!+\bs_{1,H}^{\ell,m} -\by_1^{\ell,m} \parallel^2 \medskip\\
\quad \text{s.t.} \quad  \mB_{0,0}^{(l),m} \bs_{1,H}^{\ell,m}=\mathbf{0}, \quad \mB_{1,0}^{(l),m  \, T} \bs_{1,H}^{\ell,m}=\mathbf{0}.
\end{array}
\eeq
It can be easily proved that this problem admits the following closed-form solutions \cite{barb_2020}:%,\cite{jiang2011}:
\beq
\label{eq:closed_form}
\begin{array}{lll}
\hat{\bs}_{0}^{m}=(\mB_{0,0}^{(\ell),m } \mB_{0,0}^{(\ell),m\, T})^{\dagger} \mB_{0,0}^{(\ell),m} \by_{1}^{(\ell),m}\\
\hat{\bs}_{2}^{\ell,m}=(\mB_{1,0}^{(\ell),m \, T} \mB_{1,0}^{(\ell),m})^{\dagger} \mB_{1,0}^{(\ell),m\, T} \by_{1}^{(\ell),m}\\
\hat{\bs}_{1,H}^{\ell,m}=\by_1^{(\ell),m}-
\mB_{0,0}^{(\ell),m \, T}\hat{\bs}_{0}^{\ell,m}-\mB_{1,0}^{(\ell),m}\hat{\bs}_{2}^{\ell,m}
\end{array}
\eeq
where ${}^{\dagger}$ denotes the Moore-Penrose pseudo-inverse. 
As  numerical example, we consider the two communication networks illustrated in Fig. \ref{fig:CMC_2}, where nodes represent devices  emitting  data flow packets. The two networks are connected through a set of cross-edges, with some nodes having control functionalities. Our goal is to recover from noisy observations the flows over the cross-edges  between the two networks and identify harmonic cross-hubs. Considering the cross-Laplacian matrix $\mL_{0,0}^{(\ell),m}$, we  estimate the cross-edge signals using the closed-forms in (\ref{eq:closed_form}).
\begin{figure}[!t]
\centering
\includegraphics[width=8.9cm,height=4.5cm]{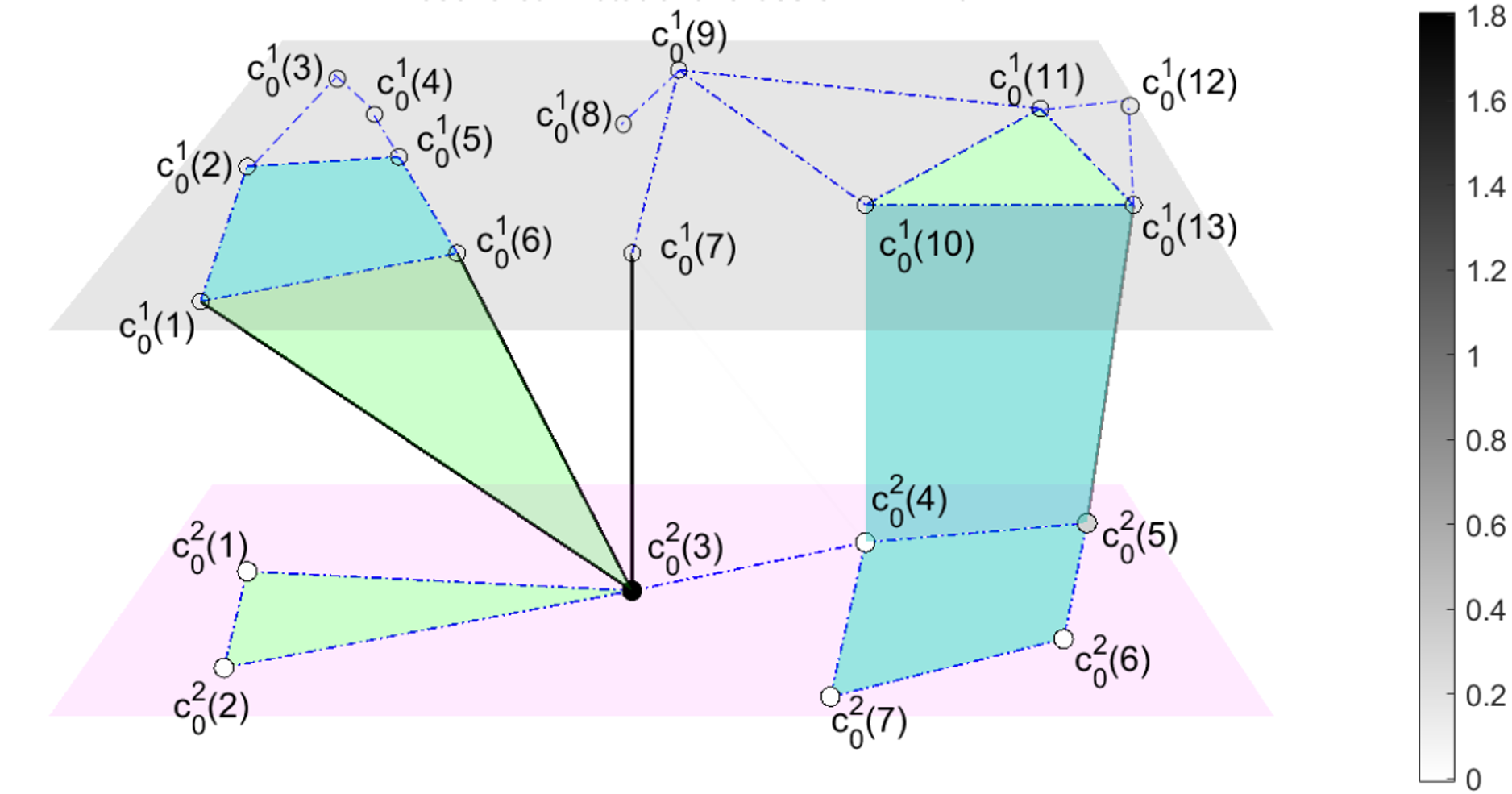}%{MC1n2.jpg}
\caption{Recovered   cross-edge and node signals.}
\label{fig:CMC_3}
\end{figure}
\begin{figure}[t]
\centering
\includegraphics[width=8.9cm,height=4.4cm]{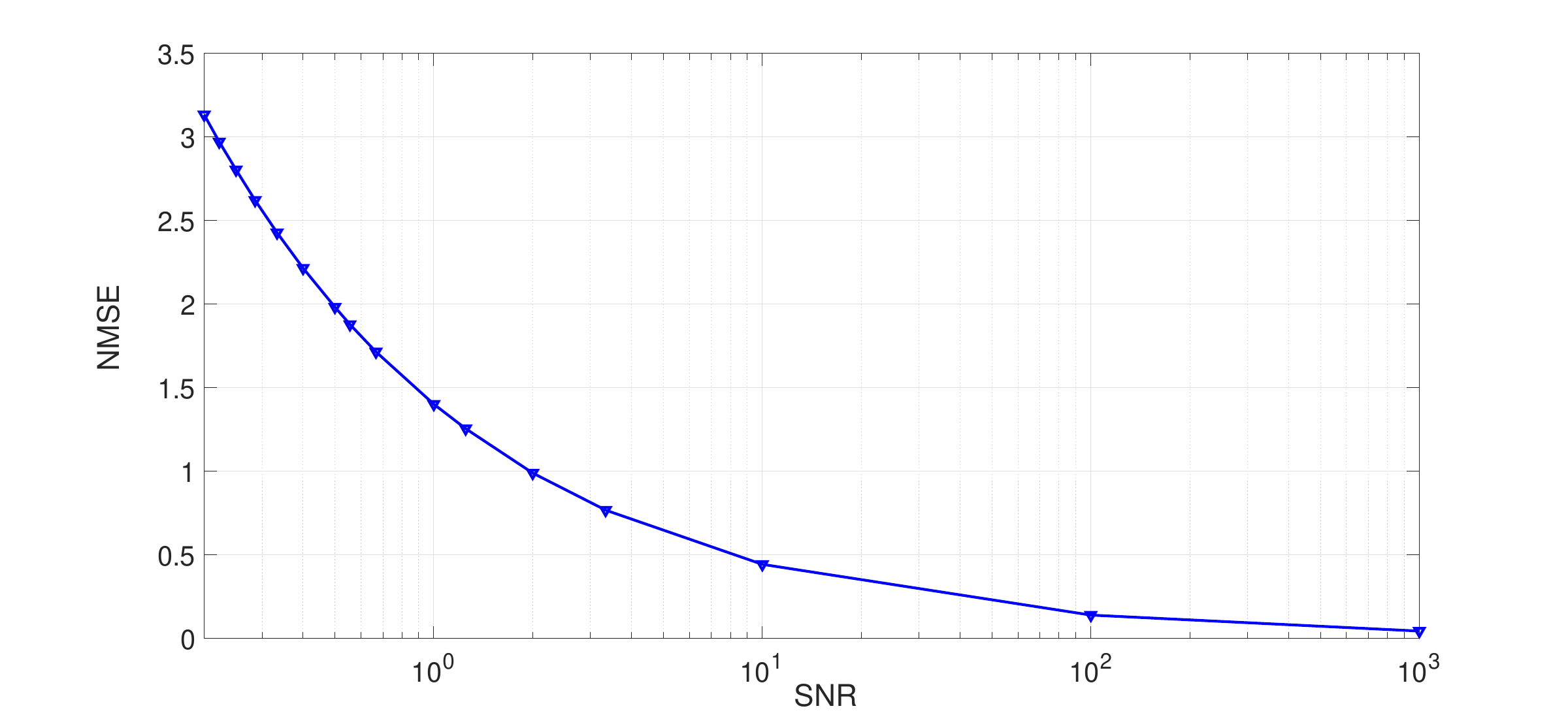}%{MC1n2.jpg}
\caption{Normalized mean squared error versus SNR.}
\label{fig:CMC_4}
\end{figure}
In Fig. \ref{fig:CMC_2} we represent the intensity of  the recovered harmonic signal $\hat{\bs}_{1, H}^{1,2}$ over the cross-edges between the two networks.  It can be observed as the harmonic signals tend to be highest on the cross-edges surrounding the two cones identified by the nodes $(c_{0}^{1}(6),c_{0}^{2}(3),c_{0}^{1}(7))$ and $(c_{0}^{1}(7),c_{0}^{2}(4),c_{0}^{1}(10))$. The two cross-hubs are the nodes $c_{0}^{2}(3)$ and $c_{0}^{2}(4)$, albeit the first cross-hub $c_{0}^{2}(3)$ has a key role in the inter-connectivity between the two networks, since it controls two clusters of nodes on the first network. Furthermore,  removing $c_{0}^{2}(3)$ the two clusters on layer $1$ are disconnected. To evaluate the activity of the cross-hubs, we derive from the estimated  edge signal $\hat{\bs}_{1}^{\ell,m}$, the  signal $\text{div}^{(\ell),m}(\hat{\bs}_{1}^{\ell,m})$.  Then, in Fig. \ref{fig:CMC_3}  we represent over the cross-edges the intensity of the estimated signals $\mB_{0,0}^{(1),2 \, T} \hat{\bs}_0^{m}$, and, on the nodes of the lower layer, the signal $\text{div}^{(\ell),m}(\hat{\bs}_{1}^{\ell,m})$. The intensities of the signals are encoded by the colors of cross-edges and nodes (in grayscale).
It can be noted as the highest node value is observed over the cross-hub $c_0^{2}(3)$. Finally,  Fig. \ref{fig:CMC_4} illustrates the average normalized squared error $\text{NMSE}:= \frac{\parallel \hat{\bs}_1^{\ell,m}-{\bs}_1^{\ell,m}\parallel}{ \parallel {\bs}_1^{\ell,m}\parallel}$ versus the signal-to-noise ratio $\text{SNR}=\sigma_1^{2}/\sigma_n^{2}$. As expected, the estimation error decreases as the SNR increases.
%\vspace{-0.2cm}
\section{Conclusions}
In this paper, we introduced the processing of signals defined on cell multicomplexes, a new representation %class
of topological spaces capable of capturing both intra- and inter-layers higher-order interactions across different networks. We showed how cross-Laplacians matrices are effective algebraic  descriptors for representing signals over CMCs. We focused on  filtering  noisy cross-edges flows, showing how to identify  harmonic cross-hubs on one layer that control the topology of other layers. Future developments will focus on extending  the proposed framework from both a theoretical and an applied perspective.

\bibliographystyle{IEEEbib}
\bibliography{reference}

\begin{thebibliography}{10}

\bibitem{boccaletti2006complex}
S.~Boccaletti, V.~Latora, Y.~Moreno, M.~Chavez, and D.-U. Hwang,
\newblock ``Complex networks: Structure and dynamics,''
\newblock {\em Physics reports}, vol. 424, no. 4-5, pp. 175--308, 2006.

\bibitem{de2013mathematical}
M.~De~Domenico, A.~Sol{\'e}-Ribalta, E.~Cozzo, M.~Kivel{\"a}, Y.~Moreno, M.~A. Porter, S.~G{\'o}mez, and A.~Arenas,
\newblock ``Mathematical formulation of multilayer networks,''
\newblock {\em Phys. Review X}, vol. 3, no. 4, pp. 041022, 2013.

\bibitem{bianconi2021higher}
G.~Bianconi,
\newblock {\em Higher-order networks},
\newblock Cambridge University Press, 2021.

\bibitem{10.1093/gigascience/gix004}
M.~De~Domenico,
\newblock ``Multilayer modeling and analysis of human brain networks,''
\newblock {\em GigaScience}, vol. 6, no. 5, pp. gix004, 02 2017.

\bibitem{breedt2023multimodal}
L.~C. Breedt, F.~A.~N. Santos, et~al.,
\newblock ``Multimodal multilayer network centrality relates to executive functioning,''
\newblock {\em Network Neuroscience}, vol. 7, no. 1, pp. 299--321, 2023.

\bibitem{liu2020robustness}
X.~Liu, E.~Maiorino, et~al.,
\newblock ``Robustness and lethality in multilayer biological molecular networks,''
\newblock {\em Nature communications}, vol. 11, no. 1, pp. 6043, 2020.

\bibitem{CRAINIC20221}
T.G. Crainic, B.~Gendron, and M.R. {Akhavan Kazemzadeh},
\newblock ``A taxonomy of multilayer network design and a survey of transportation and telecommunication applications,''
\newblock {\em European Journal of Operational Research}, vol. 303, no. 1, pp. 1--13, 2022.

\bibitem{krishnagopal2023topology}
S.~Krishnagopal and G.~Bianconi,
\newblock ``Topology and dynamics of higher-order multiplex networks,''
\newblock {\em Chaos, Solitons \& Fractals}, vol. 177, pp. 114296, 2023.

\bibitem{moutuou2023}
E.~M. Moutuou, O.~B.~K. Ali, and H.~Benali,
\newblock ``Topology and spectral interconnectivities of higher-order multilayer networks,''
\newblock {\em Frontiers in Complex Systems}, vol. 1, pp. 1281714, 2023.

\bibitem{klette2000cell}
R.~Klette,
\newblock ``Cell complexes through time,''
\newblock in {\em Vision Geometry IX}. Int. Soc. for Opt. and Photon., 2000, vol. 4117, pp. 134--145.

\bibitem{grady2010}
L.~J. Grady and J.~R. Polimeni,
\newblock {\em Discrete calculus: Applied analysis on graphs for computational science},
\newblock Sprin. Scie. \& Busin. Media, 2010.

\bibitem{sardellitti2024topological}
S.~Sardellitti and S.~Barbarossa,
\newblock ``Topological signal processing over generalized cell complexes,''
\newblock {\em IEEE Trans. Signal Process.}, 2024.

\bibitem{Lim}
L.-H. Lim,
\newblock ``{H}odge {L}aplacians on graphs,''
\newblock {\em S. Mukherjee (Ed.), Geometry and Topology in Statistical Inference, Proc. Sympos. Appl. Math., 76, AMS}, 2015.

\bibitem{barb_2020}
S.~Barbarossa and S.~Sardellitti,
\newblock ``Topological signal processing over simplicial complexes,''
\newblock {\em IEEE Trans. Signal Process.}, vol. 68, pp. 2992--3007, March 2020.

\end{thebibliography}

\end{document}